\documentclass[aps,prl,twocolumn,nofootinbib,superscriptaddress,floatfix,notitlepage]{revtex4-1}

\usepackage{graphicx}
\usepackage{amsmath,amsfonts,amssymb}
\usepackage{color}
\usepackage[breaklinks,colorlinks,urlcolor=blue,citecolor=blue,linkcolor=magenta]{hyperref}
\usepackage{verbatim}

\newcommand{\be}{\begin{equation}} 
\newcommand{\ee}{\end{equation}}
\newcommand{\bea}{\begin{equation}\begin{aligned}} 
\newcommand{\eea}{\end{aligned}\end{equation}}
\newcommand{\td}{{\rm d}}

\newcommand{\papertitle}{Impact of cosmic expansion on gravitational wave spectra \\ from strongly supercooled first-order phase transitions}

\begin{document}

\title[]{\papertitle}

\author{Marek Lewicki}
\email{marek.lewicki@fuw.edu.pl}
\affiliation{Faculty of Physics, University of Warsaw ul.\ Pasteura 5, 02-093 Warsaw, Poland}
\affiliation{Astrocent, Nicolaus Copernicus Astronomical Center Polish Academy of Sciences, ul. Rektorska 4, 00-614, Warsaw, Poland}
\author{Ville Vaskonen}
\email{ville.vaskonen@pd.infn.it}
\affiliation{Dipartimento di Fisica e Astronomia, Universit\`a degli Studi di Padova, Via Marzolo 8, 35131 Padova, Italy}
\affiliation{Istituto Nazionale di Fisica Nucleare, Sezione di Padova, Via Marzolo 8, 35131 Padova, Italy}
\affiliation{Keemilise ja Bioloogilise F\"u\"usika Instituut, R\"avala pst. 10, 10143 Tallinn, Estonia}

\begin{abstract}
We compute the gravitational wave spectra from strongly supercooled first-order phase transitions, explicitly incorporating the evolution of the background metric across the transition from thermal inflation to radiation domination. We find that the spectral shape remains largely unchanged apart from a causality-induced super-horizon tail. However, in contrast to standard expectations, for slow transitions we show that the peak amplitude and frequency exhibit a weaker dependence on the transition rate $\beta$ than the usual scaling of $\propto \beta^{-2}$ and $\propto\beta$, respectively.
\end{abstract}

\maketitle

\vspace{5pt}\noindent\textbf{Introduction --} Cosmological first-order phase transitions are among the strongest possible sources of gravitational waves (GWs) from the early Universe. The resulting GW spectrum can be probed by current LVK~\cite{LIGOScientific:2025kry} and upcoming observatories, such as ET~\cite{ET:2025xjr}, AION/AEDGE~\cite{Badurina:2019hst,AEDGE:2019nxb}, LISA~\cite{LISA:2017pwj} and SKA~\cite{Janssen:2014dka}. In particular, for strongly supercooled transitions, the shape of the GW background can be measured with very high precision~\cite{Caprini:2024hue}. It is therefore crucial to obtain accurate theoretical predictions for the GW spectrum. Furthermore, the observation of a GW background by pulsar timing arrays~\cite{NANOGrav:2023gor,EPTA:2023fyk,Reardon:2023gzh,Xu:2023wog} may have originated from an early-Universe phase transition, but only if that transition was both very strong and slow~\cite{NANOGrav:2023hvm, Gouttenoire:2023bqy, Ellis:2023oxs, Fujikura:2023lkn}.

Numerous studies have estimated the GW spectrum from phase transitions in different cases~\cite{Kosowsky:1992vn,Kamionkowski:1993fg,Huber:2008hg,Hindmarsh:2013xza,Hindmarsh:2015qta,Konstandin:2017sat,Cutting:2018tjt,Hindmarsh:2019phv,Cutting:2019zws,Jinno:2020eqg,Dahl:2021wyk,Cutting:2022zgd,Auclair:2022jod,Jinno:2022mie,RoperPol:2023dzg,Dahl:2024eup,Jinno:2024nwb,Caprini:2024gyk,Giombi:2025tkv}. In a series of papers~\cite{Lewicki:2020jiv,Lewicki:2020azd,Lewicki:2022pdb}, we have developed a hybrid approach to calculate the GW spectrum generated in strongly supercooled phase transitions. This paper continues that line of work and adds a crucial component in the GW spectrum calculation: the cosmic expansion.

Strongly supercooled phase transitions are typically realized in quasi-conformal models~\cite{Jinno:2016knw,Iso:2017uuu,Marzola:2017jzl,Prokopec:2018tnq,Marzo:2018nov,Baratella:2018pxi,VonHarling:2019rgb,Aoki:2019mlt,DelleRose:2019pgi,Wang:2020jrd,Ellis:2020nnr,Baldes:2020kam,Baldes:2021aph,Lewicki:2021xku,Levi:2022bzt,Sagunski:2023ynd,Lewicki:2024sfw}. They are characterized by a period of exponential expansion known as thermal inflation~\cite{Lyth:1995ka} and, according to previous estimates in the literature, can produce some of the strongest GW signals among cosmological phase transitions. The amplitude of the resulting GW background increases for slower transitions, as bubbles of true vacuum have more time to grow before colliding and accumulate a significant amount of energy. Including the cosmic expansion is essential for slow phase transitions that can have a duration comparable to the Hubble time.

In a radiation-dominated background, the effect of cosmic expansion on GW production has been taken into account in lattice simulations~\cite{Child:2012qg, Arapoglu:2023ljz} and in analytical treatments based on the approach of~\cite{Jinno:2016vai}, as implemented in Refs.~\cite{Zhong:2021hgo, Yamada:2025hfs}. In this work, we focus on strongly supercooled transitions and explicitly include the period of thermal inflation and the subsequent reheating driven by the phase transition.

\vspace{5pt}\noindent\textbf{Phase transition dynamics --} We consider the commonly used exponential bubble nucleation rate,
\be \label{eq:Gamma}
    \Gamma(t) = H_n^4 e^{\beta t} \,,
\ee
which can be understood as an expansion of the nucleation action to the lowest non-trivial order around the time $t=0$ when the transition starts, $\Gamma(t=0) = H(t=0)^4\equiv H_n^4$~\cite{Lewicki:2024sfw}. The probability of finding a given point in the false vacuum is~\cite{Guth:1982pn, Ellis:2018mja}
\be
    P(t) = \exp\left[-\frac{4\pi}{3} \!\int_{-\infty}^t \!\!\td t_n \Gamma(t_n) a(t_n)^3 R(t_n,t)^3 \right] \,,
\ee
where the exponent gives the expected number of bubbles that could have reached that point by time $t$. Here, $a(t)$ denotes the scale factor and $R(t_n,t)$ the bubble radius. We choose $a(t=0) = 1$. Assuming that the interactions of the bubble with the surrounding plasma and the initial bubble radius are negligible, the comoving bubble radius increases linearly with conformal time $R(t_n,t) \approx \eta(t) - \eta(t_n)$. 

We consider transitions in which the vacuum energy density $\Delta V$ of the false vacuum dominates over the radiation energy density before the transition, $H_n^2 \approx 8\pi G \Delta V/3$. The nucleation and expansion of bubbles decrease the average vacuum energy density $\rho_v(t) = P(t) \Delta V$ as the bubbles convert the vacuum energy into the energy of the bubble walls whose averaged energy density scales as radiation~\cite{Ellis:2019oqb,Lewicki:2023ioy}. The radiation energy density $\rho_r(t)$, including the energy in the bubble walls, follows the continuity equation $\partial_t \rho_r(t) + 4 H(t) \rho_r(t) = -\partial_t \rho_v(t)$, and the Hubble rate is given by the Friedmann equation $H(t)^2 = 8\pi G (\rho_r(t) + \rho_v(t))/3$. 

We solve the evolution of the vacuum and radiation energy densities numerically, together with the Hubble rate and the scale factor. Then, using that background evolution, we nucleate bubbles following the rate~\eqref{eq:Gamma} within a fixed comoving volume with periodic boundary conditions. As discussed in the following, we use this setup to compute the GW spectrum from the phase transition.

\begin{figure}
\centering
\includegraphics[width=0.9\columnwidth]{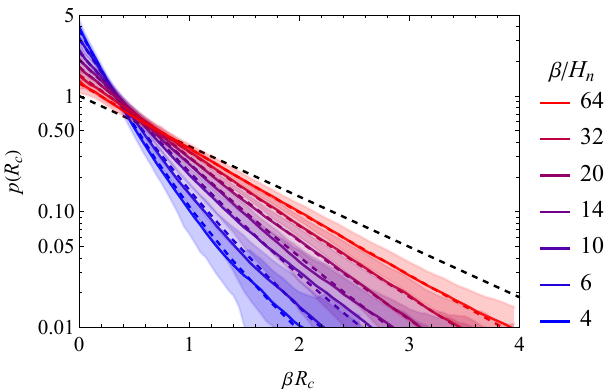}
\caption{The distributions of bubble radii at the moment of collision. The solid curves show the result obtained from numerical simulations by averaging over many realizations, with the bands indicating their spread. The dashed curves show the corresponding analytical estimate obtained using Eq.~\eqref{eq:pRc}. The black dashed line shows the Minkowski space result $p(R_c) = \beta e^{-\beta R_c}$.}
\label{fig:pRc}
\end{figure}

To conclude the discussion on phase transition dynamics, we study when the bubbles collide with other bubbles. The probability that a wall element of a bubble that nucleated at $t_n$ collided with another bubble at time $t$ in the range $t_c < t < t_c+\td t_c$ can be expressed as
\be
    \td t_c p(t_c;t_n) \propto - \td t_c a(t_n)^3 \Gamma(t_n) P(t_n + t_c) P'(t_n + t_c) \,.
\ee
Consequently, the probability distribution of the bubble radius at which a bubble surface element collides with the surface of another bubble is given by
\bea \label{eq:pRc}
    &p(R_c) \propto \int \td t_n \td t_c p(t_c,t_n) \delta(R(t_n + t_c;t_n) - R_c) \\
    &\quad = -\int {\rm d} t_n a(t_n)^3 \Gamma(t_n) a(t) P(t) P'(t) \bigg|_{t: \,R(t;t_n) = R_c} \,.
\eea
For $\beta/H_n\gg 1$, the transition is very fast compared to the cosmic expansion and, as seen in Fig.~\ref{fig:pRc}, the distribution of $R_c$ at large $\beta/H_n$ approaches the Minkowski space result $p(R_c) = \beta e^{-\beta R_c}$~\cite{Hindmarsh:2019phv,Lewicki:2022pdb}. A clear deviation from the Minkowski space result is observed for $\beta/H_n \lesssim 100$ as the $R_c$ distribution becomes steeper for slower transitions. The solid curves in Fig.~\ref{fig:pRc} and the bands surrounding them are obtained from numerical simulations.\footnote{The collision radius of a wall element in direction $\hat{x}$ is determined as
\begin{equation*} \label{eq:Rc}
    R_{b,c}(\hat{x}) = \frac12 \min_{b'} \left[\frac{|\vec{x}_{b,n} - \vec{x}_{b',n}|^2-\Delta\eta_{b,b'}^2}{\hat{x}\cdot(\vec{x}_{b,n} - \vec{x}_{b',n})+\Delta\eta_{b,b'}}\right]\,,
\end{equation*}
where $\vec{x}_{b,n}$ and $\vec{x}_{b',n}$ are the position vectors of the nucleation centers of the bubbles $b$ and $b'$ and $\Delta\eta_{b,b'} \equiv \eta_{b,n}-\eta_{b',n}$ denotes the conformal time difference between their nucleation.} An excellent agreement between the simulation results and Eq.~\eqref{eq:pRc} is evident.

\vspace{5pt}\noindent\textbf{Gravitational wave production --} 
The Einstein's equations for the metric perturbations $h_{ij}(t,\vec{x})$ read
\be \label{eq:Einstein}
    \Box h_{ij}^{\rm TT}(\vec{x}) = \frac{16\pi G}{a^2} T_{ij}^{\rm TT}(\vec{x}) \,,
\ee
where $\Box = \partial_t^2 + 3H \partial_t - a^{-2} \nabla^2$, TT denotes the transverse traceless projection and $T_{ij}$ is the stress-energy tensor of the perturbations that source the GWs. In the Fourier space, we write the perturbed Einstein's equation~\eqref{eq:Einstein} as
\be \label{eq:u}
    \left[\partial_t^2 + 3H \partial_t + \frac{k^2}{a^2}\right]\! u_{ij}(\vec{k}) = \Lambda_{ij,lm}(\hat{k}) \frac{\tilde{T}_{ij}(\vec{k})}{a^2 \Delta V} \,,
\ee
where $\Lambda_{ij,lm}(\hat{k})$ is the TT projection operator and we have introduced an auxiliary tensor $u_{ij}$ defined via\footnote{Notice that the mass dimension of $u_{ij}$ is $-5$.}
\be
    \tilde h_{ij}^{\rm TT}(t,\vec{k}) = 16\pi G \Delta V u_{ij}(t,\vec{k}) \,.
\ee

The stress-energy tensor of the scalar field $\phi$ associated with the phase transition, excluding the pure trace part that does not source GWs, reads 
\be
    T_{ij}(\vec{x}) = \partial_i \phi \partial_j \phi \,.
\ee
We work in the thin-wall limit where the overlap of the walls can be neglected. In this limit, the Fourier transform of the stress-energy tensor can be expressed, using the spherical symmetry of the bubbles, as~\cite{Kosowsky:1992vn} 
\be \label{eq:Ttensor}
     \frac{\tilde{T}_{ij}(\vec{k})}{a^2 \Delta V} = \sum_b e^{-i \vec{k}\cdot\vec{x}_{b,n}} \!\int \!\td \Omega_{\hat x} \,e^{-i \vec{k}\cdot \hat{x} R_b} \hat{x}_i \hat{x}_j \! \int \! \td r\, r^2 \frac{(\partial_r \phi)^2}{a^2 \Delta V} \,,
\ee
where the sum runs over all bubbles in a given volume, and $\vec{x}_{b,n}$ and $R_b$ denote the nucleation center positions and the bubble radii.\footnote{We use the convention where the Fourier transform of $X(t,\vec{x})$ is $\tilde{X}(t,\vec{k}) = \int \td^3 x \,e^{-i \vec{k}\cdot\vec{x}} X(t,\vec{x})$.}

The radial integral in~\eqref{eq:Ttensor} can be estimated by analyzing a single bubble. Inside the bubble, the potential energy is negligible, and at the wall the kinetic and gradient energies quickly equilibrate: $(\partial_t \phi)^2 = (\partial_r \phi)^2/a^2$, as the wall expands along the trajectory $\partial_\eta R_b = a^{-1} \partial_t R_b = 1$. So, the energy of the bubble in a differential solid angle $\td \Omega_{\hat x}$ is $\td E_{\rm b}/\td \Omega_{\hat x} = a^3 \int \td r\,r^2/a^2 (\partial_r \phi)^2$. In the thin-wall approximation, this energy is also given by $\td E_{\rm b}/\td \Omega_{\hat x} = (a R_b)^2 \gamma_b \sigma$, where $\sigma$ is the wall tension and $\gamma_b$ the Lorentz factor of the wall. Finally, we find
\bea \label{eq:rint}
    \int \!\td r\, r^2 \frac{(\partial_r \phi)^2}{a^2 \Delta V} &= \frac{\gamma_b \sigma}{a \Delta V} R_b^2 \\
    &\approx \frac{R_b^3}{3}
    \begin{cases}
        1\,, & R_b \leq R_{b,c}\\
        \frac{a_c}{a}\left[\frac{R_{b,c}}{R_b}\right]^3\,, & R_b > R_{b,c}
    \end{cases} .
\eea
For the the second line we have used the solution of the equation of motion~\cite{Lewicki:2023ioy}
\be \label{eq:gamma}
    \frac{\td \gamma_b}{\td R_b} + \frac{2 \gamma_b}{R_b} \left[ 1 + \frac32 a H R_b \right] = \begin{cases}
        a \frac{\Delta V}{\sigma} \,, & R_b \leq R_{b,c} \\
        0 \,, & R_b > R_{b,c}
    \end{cases}\,,
\ee
where $R_{b,c}$ denotes the radius at which the bubble wall element collides with another bubble, in the limit $a H R_b \ll 1$.\footnote{We have dropped the factor $\sqrt{1-1/\gamma_b^2}$ from the second term inside the brackets, as, in the absence of friction, $\gamma_b\gg 1$ in the regime where $aHR_b$ can not be neglected.} In our numerical computations, we use the complete numerical solution of Eq.~\eqref{eq:gamma}. 

Neglecting cosmic expansion, Eq.~\eqref{eq:rint} corresponds to the bulk flow model~\cite{Konstandin:2017sat}, while setting the case $R_b > R_{b,c}$ to zero gives the envelope approximation~\cite{Kosowsky:1992vn}. Furthermore, we allow for additional energy dissipation of the scalar field shells after the collisions by multiplying the case $R_b > R_{b,c}$ by $R_{b,c}/R_b$. We call this the \emph{dissipative bulk flow model}. Such dissipation is seen in simulations in the Minkowski background~\cite{Lewicki:2020azd} if the scalar field is real or it is associated with a gauged U$(1)$ symmetry. It is also seen in simulations of transitions where the energy is transferred to fluid shells rather than the acceleration of the bubble wall~\cite{Lewicki:2022pdb}.

We solve the perturbed Einstein's equation, Eq.~\eqref{eq:u}, numerically.\footnote{The GW equation is often solved using the Green's function (see e.g.~\cite{Kosowsky:1992vn,Guo:2020grp}). This is particularly convenient in Minkowski space and in the radiation dominated FLRW universe. In general FLRW universes the Green's functions, however, become more complicated~\cite{Caldwell:1993xw}.} The resulting total GW energy density reads
\bea
    \rho_{\rm GW}(t) &= \frac{1}{32\pi G} \sum_{i,j} \left\langle |\partial_t h_{ij}^{\rm TT}(\vec{x})|^2 \right\rangle \\
    &= \frac{\Delta V^2 G}{\pi^2 L^3} \sum_{i,j} \int \td^3 k \big| \partial_t u_{ij}\big|^2 \,,
\eea
where $\langle \cdot \rangle$ denotes spatial average. For the second line, we have used $\langle |f(\vec{x})|^2 \rangle = (2\pi L)^{-3} \int \td^3 k |\tilde f(\vec{k})|^2$ where $L^3$ denotes the comoving volume over which the average is taken, that is, the simulation volume.

After some time $t_{\rm end}$ the transition ends and the GW source becomes sufficiently diluted. From then on, GWs propagate freely in a radiation-dominated background $(\partial_\tau^2 + k^2) a u_{ij} \approx 0$, and the solution can be trivially extended beyond $t_{\rm end}$. We then average the spectrum $\td \rho_{\rm GW}/\td \ln k$ over one period and take the limit $t\gg t_{\rm end}$. The final GW spectrum is given by
\be \label{eq:OmegaGW}
    \Omega_{\rm GW}(k) \equiv \frac{1}{\rho_{\rm tot}}\frac{\td \rho_{\rm GW}}{\td \ln k} = \left(\frac{H_n}{\beta}\right)^2 S(k) \,,
\ee
where $\rho_{\rm tot} = 3H^2/(8\pi G)$, and the function $S(k)$ incorporates the shape of the spectrum:
\be \label{eq:S}
    S(k) \!=\! \Theta k^3 \!\int\! \td^3 \hat{k} \sum_{i,j} \! \left[ \bigg|u_{ij} + \frac{\partial_t u_{ij}}{H}\bigg|^2 + \bigg|\frac{k}{aH} u_{ij}\bigg|^2 \right]\!\bigg|_{t_{\rm end}} ,
\ee
with $\Theta = 3\beta^2 H_n^2/(16 \pi^3 L^3)$. Conventionally, we have factored out $(H_n/\beta)^2$ in Eq.~\eqref{eq:OmegaGW}. In a Minkowski background, this prefactor captures the full dependence of the GW spectrum on $\beta$.

\vspace{5pt}\noindent\textbf{Results --} We consider a range of values of $\beta/H_n$ and perform 30 numerical simulations of the phase transition for each case. In these simulations, we nucleate and evolve bubbles, compute the stress energy tensor~\eqref{eq:Ttensor}, solve the perturbed Einstein's equation~\eqref{eq:u} using the fourth-order Runge-Kutta method for 50 wavenumbers between $k_{\rm min} = 2\pi/(4L)$, where $L$ denotes the simulation box size and $k_{\rm max} = 20\beta$, and compute the GW spectrum by evaluating Eq.~\eqref{eq:S}.

\begin{figure}
\centering
\includegraphics[width=\columnwidth]{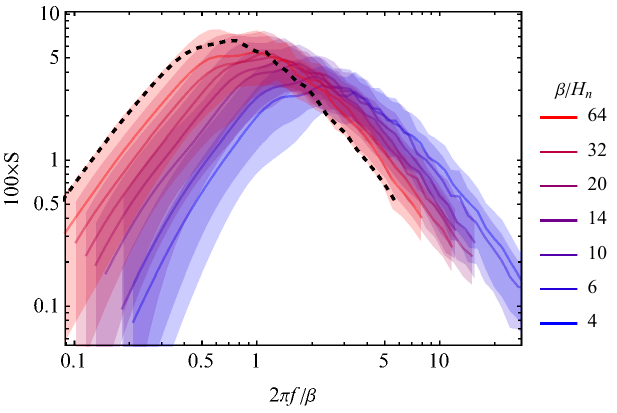}
\caption{The simulation results for the GW spectra in the dissipative bulk flow model, obtained by averaging the realisations in each frequency bin. The bands around each curve indicate the spread of the results. The black dashed curve shows the result obtained by neglecting the cosmic expansion.}
\label{fig:Spectra}
\end{figure}

Fig~\ref{fig:Spectra} shows the resulting GW spectra in the dissipative bulk flow model as averages over all the realisations, with the bands indicating their spread. The spectra are already scaled with the factor $(\beta/H_n)^{-2}$ as in Eq.~\eqref{eq:OmegaGW}. However, a clear trend as a function of $\beta/H_n$ is still visible: The spectra move towards higher frequencies and lower amplitudes with decreasing $\beta/H_n$. At large $\beta/H_n$ the spectrum lies near the result obtained by neglecting the cosmic expansion, shown in black. This is expected as for a large $\beta/H_n$ the transition lasts for a very short fraction of a Hubble time.

To extract the shape of the spectra as a function of the frequency $f = k/2\pi$, we fit the simulation results using a log-normal ansatz
\be
    p(S|f) = \frac{1}{\sqrt{2\pi}\sigma} \exp\!\left[-\frac{(\ln S - \ln \bar{S}(f))^2 }{2\sigma^2} \right] \,,
\ee
whose mean follows a broken power-law
\be
    \bar{S}(f) = \frac{A(a+b)^c}{\left[ b\left(\frac{f}{f_p}\right)^{-\frac{a}{c}} + a\left(\frac{f}{f_p}\right)^{\frac{b}{c}} \right]^{c}} \,.
\ee
The variance of the log-normal ansatz describes the scatter between the simulations. We perform a Markov chain Monte Carlo (MCMC) inference of the model parameters $\vec{\theta} = (A,f_p,a,b,c,\sigma)$ by sampling the likelihood 
\be
    \ln \mathcal{L}(\vec{\theta}) = \sum_{j,i} p(S_{i,j}|f_j,\vec{\theta}) \,,
\ee
where $j$ labels the frequencies and $i$ the simulations. 

\begin{figure}
\centering
\includegraphics[width=\columnwidth]{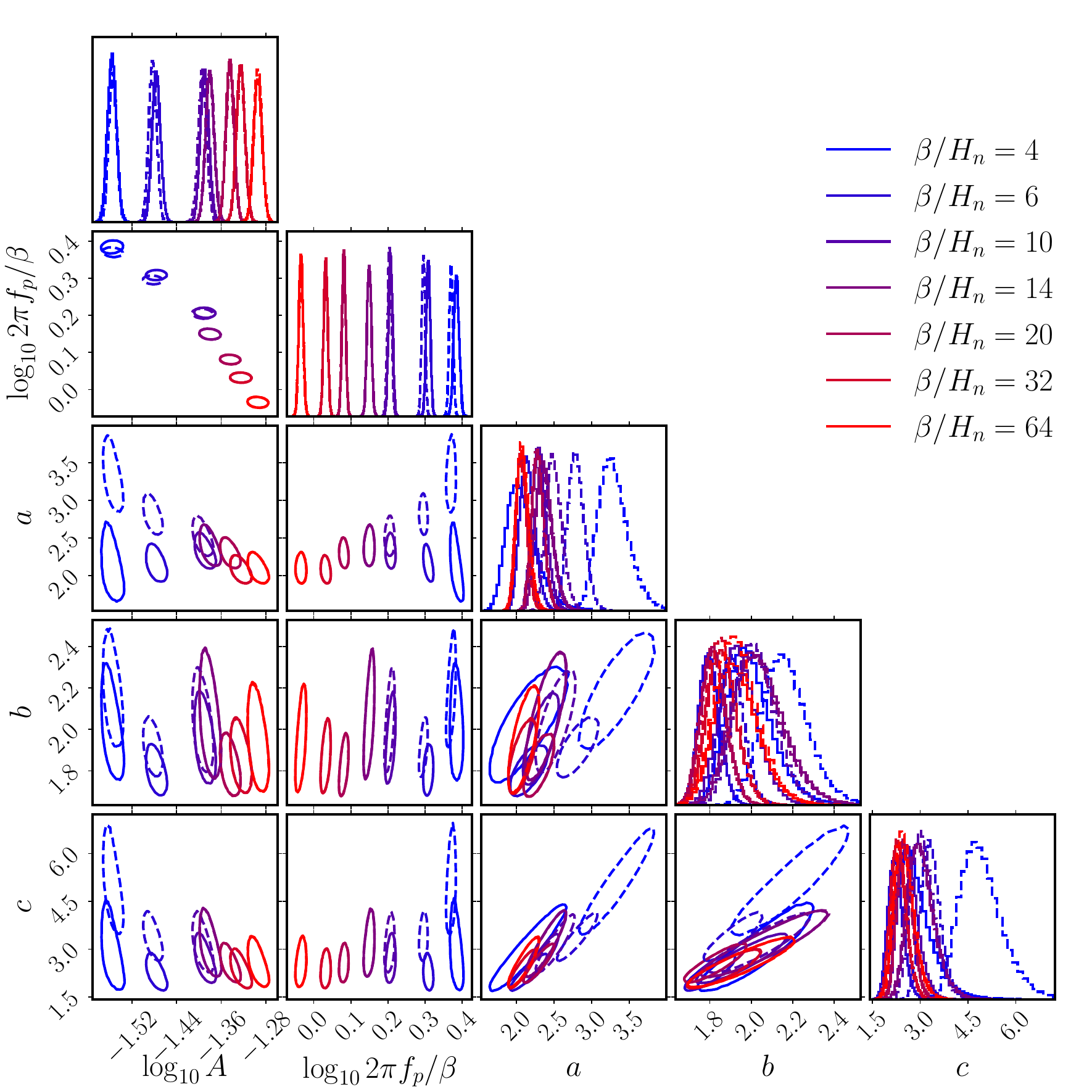}
\caption{The $95\%$ CL regions of the posteriors of the fit of the simulation results in the dissipative bulk flow model. Dashed lines are obtained using a simple broken power-law fit, while solid lines show the results with a double broken power law featuring a causality tail below the horizon radius scale (see Eq.~\eqref{eq:fH}).}
\label{fig:posteriors}
\end{figure}

The $95\%$ confidence level (CL) posteriors of the inference marginalized over $\sigma$ are shown by the solid contours in Fig.~\ref{fig:posteriors} for the dissipative bulk flow model. For large $\beta/H_n$, the low- and high-frequency slopes are around $a \approx b \approx 2$, consistent with the results we find by neglecting the cosmic expansion. For $\beta/H_n<10$ there is a clear deviation towards steeper power-laws. In particular, for $\beta/H_n = 4$ we find that $a$ is consistent with $a\approx 3$, as expected for the tail arising from causality. 

To investigate whether inclusion of the causality driven tail would change the fit, we consider a sharp transition to $\propto f^3$ -tail below the inverse horizon radius scale, 
\be \label{eq:fH}
    f_H = \frac{a_p H_p}{2\pi}  \,,
\ee
where the subscript $p$ refers to the percolation time defined as the moment when the false vacuum fraction is $P(t_p) = 1/e$. For example, for $\beta/H_n = 4$, this frequency is $2\pi f_H/\beta \approx 0.46$, which is well in the range of our simulation results. The posteriors in this case are shown by the dashed contours in Fig.~\ref{fig:posteriors}. We see that, independent of $\beta$, the spectrum in the dissipative bulk flow are consistent with $a\approx b\approx c \approx 2$. In the same way, in the bulk flow model, we find $a\approx c \approx 1$ and $b\approx 2$.

\begin{figure}
\centering
\includegraphics[width=0.96\columnwidth]{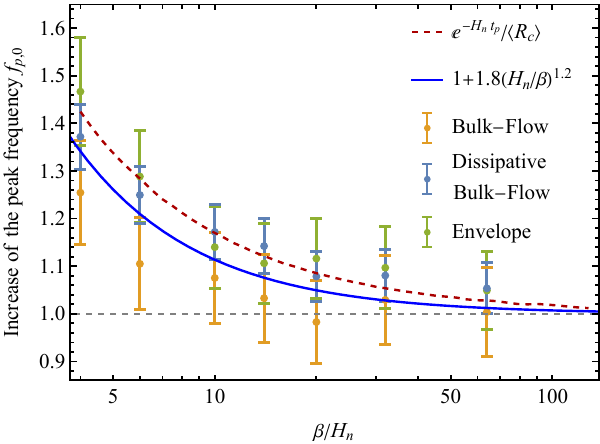}
\includegraphics[width=0.96\columnwidth]{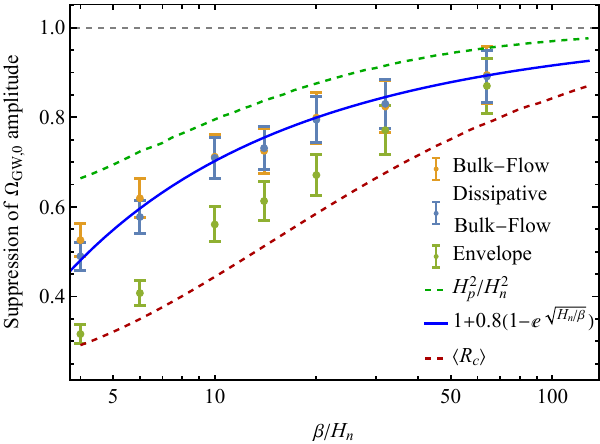}
\caption{Increase of the the peak frequency (upper panel) and suppression of the peak amplitude (lower panel) of the GW spectrum due to cosmic expansion for slow phase transitions. The data are obtained from the MCMC inference of the broken power-law fit, scaled with the results obtained neglecting the cosmic expansion.}
\label{fig:Amplitude}
\end{figure}

In order to obtain the present GW spectrum, we account for the scaling of the GW energy density as $a^{-4}$ and frequency as $a^{-1}$. We define the reheating temperature using the instantaneous reheating approximation where the transition to radiation dominance occurs abruptly at the percolation time with $a_p = e^{t_p H_n}$ and $H_p = H_n$. The reheating temperature in this approximation is $T_{\rm rh} = (30 \Delta V/\pi^2 g_*)^{1/4}$, and the GW spectrum amplitude and frequency today can be expressed as
\be
    A_0 \approx \frac{1.6\times 10^{-5}}{h^2} \left[\frac{g_*}{100}\right] \left[\frac{g_{*s}}{100}\right]^{-\frac43} A
\ee
and
\be
    f_{0} 
    \approx 2.6\!\times\! 10^{-8}\,{\rm Hz}\, \frac{T_{\rm rh}}{{\rm GeV}} \left[\frac{g_{*}}{100}\right]^{\frac12} \left[\frac{g_{*s}}{100}\right]^{-\frac13} \frac{2\pi f}{a_p H_p} \,.
\ee
Here $g_*$ and $g_{*s}$ denote the effective numbers of energy and entropy degrees of freedom at $T = T_{\rm rh}$. We find that the frequency below which the causality tail appears can be approximated by
\be \label{eq:fHfit}
    \frac{2\pi f_H(\beta)}{a_p H_p} \approx 
    1 - 0.6 \left(\frac{\beta}{H_n}\right)^{-0.67} \,,
\ee
which corresponds to the ratio of $a_p H_p$ obtained from a full numerical evaluation to that in the instantaneous reheating approximation.

The most important impact of the cosmic expansion is a decrease in the amplitude of the GW spectrum and an increase in its peak frequency for slow transitions. We show these effects in Fig.~\ref{fig:Amplitude}. The data are obtained from the MCMC inference, scaled relative to the case without cosmic expansion. In the upper panel, we see that the increase in the peak frequency of the spectrum follows the mean collision radius of the bubbles, obtained from the distribution~\eqref{eq:pRc}, scaled to the percolation time. The modification of the peak frequency today in both the bulk flow and the dissipative bulk flow models can be approximated as
\be \label{eq:fpfit}
    \frac{2\pi f_p(\beta)}{a_p H_p} \approx 0.7 \frac{\beta}{H_n} \left[ 1 + 1.8 \left(\frac{\beta}{H_n}\right)^{-1.2} \right] \,.
\ee
In the lower panel, the green dashed curve shows $(H(t_p)/H_n)^2$. We find this factor if in Eq.~\eqref{eq:OmegaGW} we factored $(\alpha/(1+\alpha))^2 (H(t_p)^2/\beta)^2$ with $\alpha = \Delta V/(\rho_{\rm tot}(T_p) - \Delta V)$ instead of $(H_n/\beta)^2$. We see that $(H(t_p)/H_n)^2$ does not fully explain the suppression in the amplitude. Some suppression may originate from the change in the bubble collision radius distribution. However, comparing with the solid dashed curve, we see that the mean bubble collision radius overestimates the suppression. For both the bulk flow and the dissipative bulk flow, the dependence of the amplitude on $\beta$ can be approximated as
\be \label{eq:Afit}
    A(\beta) \approx 0.06 \left[1 + 0.8 \left(1-e^{\sqrt{H_n/\beta}} \right) \right] \,.
\ee
For large $\beta/H_n$ our results agree with those found in earlier works~\cite{Konstandin:2017sat,Lewicki:2020jiv,Lewicki:2020azd,Lewicki:2022pdb} that neglected the effects of cosmic expansion.

\vspace{5pt}\noindent\textbf{Conclusions --} We have studied GW production in slow first-order phase transitions, taking into account the effects of cosmic expansion. We have shown that this modifies the relation between the transition rate and the resulting GW spectrum. In particular, while the overall shape of the spectrum remains largely independent of the inverse duration parameter $\beta$, the peak amplitude decreases more slowly than $\beta^{-2}$ and the peak frequency increases more slowly than $\beta$. We have provided simple approximations for these dependencies in Eqs.~\eqref{eq:fpfit} and~\eqref{eq:Afit}. Apart from the causality tail emerging at scales beyond the horizon, at the frequency given in Eq.~\eqref{eq:fHfit}, we have shown that the spectrum can be described by a broken power-law as in Eq.~\eqref{eq:S} with $a\approx b\approx c \approx 2$ for the dissipative bulk flow model and $a\approx c\approx 1$ and $b\approx 2$ for the bulk flow model.

Our results highlight the importance of including the cosmological expansion when modeling GW signals from long-lasting transitions. They provide an improved theoretical basis for interpreting the current and future GW observations in terms of underlying microphysical parameters and for distinguishing slow transitions from the conventional fast-transition regime.

\begin{acknowledgments}
\emph{Acknowledgments --} The work of M.L. was supported by  TEAMING grant Astrocent Plus (GA: 101137080) funded by the European Union, the Polish National Science Center grant 2023/50/E/ST2/00177, and the Polish National Agency for Academic Exchange within Polish Returns Programme under agreement PPN/PPO/2020/1/00013/U/00001. The work of V.V. was supported by the European Union's Horizon Europe research and innovation program under the Marie Sk\l{}odowska-Curie grant agreement No. 101065736, and by the Estonian Research Council grant RVTT7 and the Center of Excellence program TK202.
\end{acknowledgments}

\vspace{-2mm}

\bibliography{gw}

\end{document}